# Breaking the 2-loop barrier for generalized IBP reduction

A.A. Radionov and F.V. Tkachov
Institute for Nuclear Research of Russian Academy of Sciences, Moscow

**Abstract**

We discuss the problem of constructing differential operators for the generalized IBP reduction algorithms at the 2-loop level. A deeply optimized software allows one to efficiently construct such operators for the first non-degenerate 2-loop cases. The most efficient approach is found to be via the so-called partial operators that are much simpler than the complete ones, and that affect the power of only one of the polynomials in the product.

## Introduction

The IBP reduction algorithms based on the so-called integration-by-parts identities [1] are a standard tool of multiloop calculations. In ref. [2], a rather general version was suggested, based on a theorem due to Bernstein [3]. However, finding the corresponding differential operators in a systematic fashion proved to be extremely hard. In calculations, only the explicit solution for the one-loop case found in [2] was used (the so-called minimal BT-approach [4]), and some partial results related to 2-loop integrals were published [5].
We have managed to cross the 2-loop barrier for the construction of complete operators of this kind. This does not mean one can freely construct such operators for arbitrary 2-loop diagrams yet, but that one can at least explore the specifics of the problem for true non-degenerate 2-loop integrals.

## On inhibiting assumptions

Since the construction of required identities proves to be extremely cumbersome, one should turn to what can be called “the rule of vertical transcendence of interdisciplinary boundaries”: for best results, the boundaries between different levels of the solution must be transparent to the problem solver, because the obstacles that emerge at one level may have optimal solutions at another. This rule is essentially a generalization of the well-documented experience of developing complex software systems [6]. In our case, the levels are as follows:

1) The application level: the understanding of the problem, its formulation, the setup of the corresponding mathematical problems.

2) At the level of mathematics, there is a theoretical part (theorems of existence, etc.) and an implementation part (algorithms, numerical methods, etc.).

3) At the level of programming there is a sublevel of architecture (data structures, the structure of the software system) and a sublevel of coding.

One must keep in view all the above levels in the very cumbersome problem under consideration: to make choices at the level, say, mathematics with a complete understanding of the level of programming, or vice versa, because the inhibitions resulting from implicit assumptions cannot be tolerated at any level.

For example, at the application level one can hear that "only analytical answers will survive in eternity". But significant digits — the final purpose of a computation — are no less enduring. Moreover, analytical answers may be many pages long and expressed in terms of special functions that have yet to be converted to significant digits. Also, some important cases require a piece of code to be built into a Monte Carlo event generator rather than an analytical formula.

There are inhibitions at the level of problem formulation. The reference point here is that one should be able, in the final respect, to produce a numeric answer for an amplitude for given values of masses and momenta. Anything on top of that is a nice but optional bonus.

There are inhibitions due to implicit assumptions at the level of mathematics. Widely spread is a sort of belief in the magic of sophisticated mathematics (Gröbner bases etc.) implemented in powerful computer algebra systems (CAS). However, the sophistication may be a matter of linguistics, and the power carries an overhead due to the semantic gap between the domain specific language and the machine language.

There are inhibitions at the level of programming This is primarily a belief in the power of industrial-strength tools (C++, Java, python ...) that actually carry a huge burden of excessive complexity due to the pile-up of patches to correct design errors. There is also an expectation that a CAS would do all the dirty work for one, although a naпve programmer would often simply run into an intermediate blow-up.

**Generalized IBP algorithms**

The IBP reduction algorithms were discovered long ago [1] and are a standard tool of multiloop calculations. They use special identities to express more complex integrals in terms of simpler ones. One aspect of such algorithms is of interest to us here. In terms of Feynman parametrization (integrals over a standard simplex of $L-1$ dimensions where $L$ is the number of lines), the IBP identities connect integrals in such a way that some integrals run over parts of the boundary of the integration region for other integrals. This pattern emerges in the generalized IBP algorithms as well.

In the 1950's, Gelfand et al. introduced [8] analytic continuation to negative non-integer $\mu$ for expressions of the form $X^{\mu}$ treated as generalized functions, via integration by parts. For dimensional regularization (which is related to such analytic continuation, cf. [9]), the integration by parts was used for analytic continuation in [10].

In 1954, Gelfand proposed a hypothesis that for any polynomial $P(x_0, x_1, \ldots)$ of any number of variables $x_0, x_1, \ldots$ the complex power $P^\mu(x_0, x_1, \ldots)$ exists as a generalized function via analytic continuation in $\mu$. Subsequently Bernstein proved [3] that for any such $P$ there exists a finite-order differential operator $D(\partial_0, \partial_1, \ldots)$, where $\partial_i$ is partial derivative with respect to $x_i$, such that:

$$D(\partial_0, \partial_1, \ldots) P^\mu(x_0, x_1, \ldots) = b(\mu) P^{\mu-1}(x_0, x_1, \ldots), \quad b(\mu) \neq 0 \tag{1}$$

One can say that the operator $D$ decreases the power of $P$ by one. The coefficients of $D$ are polynomials of $x_i, \mu$. The essence of Bernstein's proof is that if one takes polynomials $P$ and $b$ of increasing degrees with unknown coefficients, then, after reducing the identity to a homogeneous linear system, the number of unknowns grows faster than the number of equations, whereas the differentiations ensure a mixing of monomials of different original degrees with respect to $x$, so that with sufficiently large degrees of $P$ with respect to $x_i, \mu, \partial_i$, the system saturates and begins to have non-zero solutions.

For applications to calculation of multiloop integrals one needs a (trivial) generalization of eq. (1) to the case of several polynomials $P_i$:

$$D(\partial_0, \partial_1, \ldots) \prod_i P_i^{\mu_i}(x_0, x_1, \ldots) = b(\mu) \prod_i P_i^{\mu_i - 1}(x_0, x_1, \ldots), \quad b(\mu) \neq 0 \tag{2}$$

Standard integrals with two or more loops at $d \neq 4$ have two such polynomials: one that depends only on the topology of the corresponding Feynman diagram and another one that is a linear combination of dimensional kinematical parameters, i.e. scalar products of external momenta, masses squared, etc. (In non-scalar models integrands contain yet another polynomial as a pre-factor for the product but its presence affects neither the identities (2), nor the principle of their use.)

The generalized IBP reduction consists in applying the formula several times, replacing the product of two polynomials raised to complex powers by the left hand side of eq. (2) and getting rid of derivatives via integration by parts. As a result, the powers of the polynomials are raised to values that are safe for numerical integration, whereas the boundary terms that are generated by the integration by parts, will be integrals over parts of the boundary of the original simplex, as in the case of the conventional IBP algorithms.

**The main problem**

For interesting applications to perturbative quantum field theory it is necessary to find ways to construct the operator $D$ for a given pair of polynomials $P$. Since there is only an existence theorem of unspecific nature for $D$, it is natural to search for a solution using the method of undetermined coefficients: first one makes an assumption about the degrees of $D$ with respect to $\partial_i$ and $x_i$, then $D$ is written out with undetermined coefficients, finally eq. (2) is reduced to a system of homogeneous linear equations for the undetermined coefficients. Bernsteins's result guarantees that, starting from sufficiently large degrees of $D$, the system will have non-zero solutions with $b(\mu) \neq 0$.

For one-loop integrals, the topological polynomial degenerates into 1, whereas the kinematical one is quadratic in $x$, and the solution for $D$ can be written out in an explicit form [2]. This special case proved to be useful for some 2-loop calculations, see [4].

**Example from GRACE**

The GRACE collaboration presented [5] first non-trivial examples related to 2-loop integrals, but only with one, kinematic polynomial:

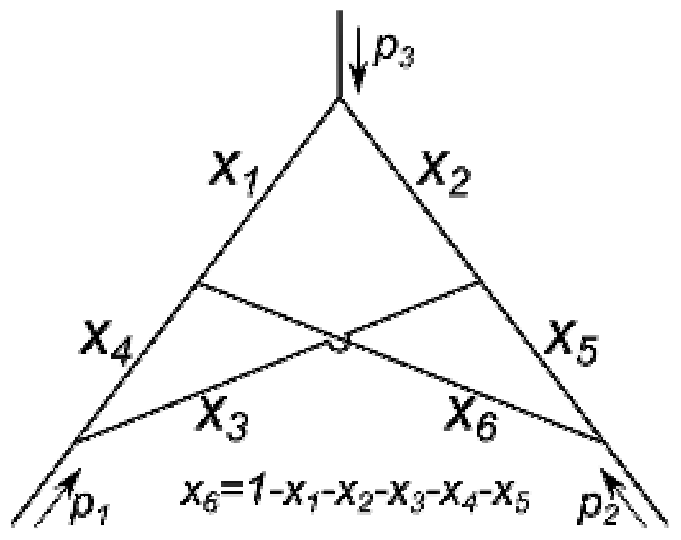

$$\int_0^1 dx_1 \ldots dx_6 \delta\left(1-\sum x\right) \frac{1}{\left(P+i\varepsilon\right)^2} \tag{3}$$

$$P = f_1 s_1 + f_2 s_2 + f_3 s_3 - f_0 \sum m_i^2 x_i$$

where

$$f_0 = (x_1+x_2)(x_3+x_4+x_5+x_6)+(x_3+x_4)(x_5+x_6)$$

$$f_1 = (x_1+x_2+x_5+x_6)x_3x_4 + x_1x_3x_6 + x_2x_4x_5$$

$$f_2 = (x_1+x_2+x_3+x_4)x_5x_6 + x_2x_3x_6 + x_1x_4x_5$$

$$f_3 = (x_3+x_4+x_5+x_6)x_1x_2 + x_2x_4x_6 + x_1x_3x_5$$

$$s_1 = 10,\ s_2 = 0,\ s_3 = 0,\ m_i^2 = i$$

If one cancels the cumbersome common factor in the results of [5], then $D$ takes the following simple form (note that it is much easier to verify such identities in any general-purpose CAS than to construct them):

$$\begin{aligned} D = & -\frac{1}{20}\partial_1^2 + \frac{1}{10}\partial_1\partial_2 - \frac{1}{20}\partial_1\partial_3 + \frac{1}{20}\partial_1\partial_4 + \frac{13}{20}\partial_1\partial_5 - \frac{1}{20}\partial_2^2 + \frac{1}{20}\partial_2\partial_3 - \frac{1}{20}\partial_2\partial_4 + \frac{7}{20}\partial_2\partial_5 \\ & -\frac{1}{20}\partial_3^2 + \frac{1}{10}\partial_3\partial_4 + \frac{13}{20}\partial_3\partial_5 - \frac{1}{20}\partial_4^2 + \frac{7}{20}\partial_4\partial_5 + 6\partial_5 - \frac{603}{20}\partial_5^2 + 3\mu\partial_5 - 3x_1\partial_2\partial_5 + 57x_1\partial_5^2 \\ & -3x_2\partial_2\partial_5 + 24x_2\partial_5^2 - 3x_3\partial_3\partial_5 + 18x_3\partial_5^2 - 3x_4\partial_3\partial_5 + 45x_4\partial_5^2 + 3x_5\partial_5^2 \end{aligned} \tag{4}$$

The operator $D$ turned out to be rather simple here. However, the complexity of construction of $D$ for two polynomials is comparable to that for the case of one polynomial which is equal to the product of the two, so that the complexity of the construction greatly increases if the topological polynomial is included into the picture, that has a second degree in the 2-loop case.

We have found a calculational scheme that allows one to construct the operators $D$ quite efficiently at least for the simpler of the non-degenerate 2-loop cases. Below we outline the scheme bottom-up, starting from the coding level.

**The coding level**

Our choice of implementation language is a popular dialect of the Oberon family [11] called Component Pascal [12]. This calls for some comments in addition to the earlier accounts [13], because the inhibitions due to the mainstream obfuscation around programming languages are huge (it is fair to say, disastrous), and the Oberon message of simplicity and minimalism confronts the mainstream views of programming at a deep level.

Oberon is an ultimate refinement of Pascal via Modula-2 by the only Turing Award winner who won the prize for language design. It is a minimalistic lucid general-purpose imperative language that presents a meticulously designed concert of basic constructs (including a polymorphic record extension which is a generic prerequisite for object-oriented programming) allowing an instantaneous single-pass modular compilation into a clean safe machine code. The essential strict static type safety, including strict type control across module boundaries, is extended to dynamic records in such a way as to enable automatic garbage collection with no overhead for the old school fortran-type programs. The type-safe dynamic (un)loading of separately compiled modules provides for a flexible incremental development cycle with an interactive look-and-feel, and a superb protection without getting in the programmer's way and eliminating the need for extraneous tools, ensuring a smooth and segviol-free design, growth and evolution of programs without loss of run-time efficiency.

This explains why Oberon's position among imperative programming languages is compared to that of the decimal positional notation among the multitude of bizarre numeric notations invented by humankind [14], and why Oberon is sometimes called Ultra Pascal from "nec plus ultra"; cf. the testimony of a Golang co-creator [15].

The net effect is that there is a growing number of examples (usually involving independent small teams tackling demanding applications), where switching to Oberon resulted in an order of magnitude or greater increase in programmers' productivity [16], thus refuting Brook's famous mantra [17]. In particular, the Oberon world is free from both the agony of obscure compilation errors and the curse of memory faults that are painfully familiar to C/C++ programmers. One cannot overestimate such a safety, achieved without compromising the efficiency of a true compiled language, in large-scale computer algebra problems.

**The level of computer algebra**

We have been using a minimalistic CAS engine that evolved from the framework BEAR v.2 which was successfully used in the rather cumbersome calculations for b-quark decays [18]. In the new version 3 the interfaces and the core algorithms changed enough to justify a renaming; the engine is now called Gulo (Latin for wolverine). It is still a very small framework, designed so as to not limit the programmer in regard of either data representation or expression size, and to allow them to control memory and files allocations to any desired degree of detail, while hiding such details in the default case. For the problems we discuss here, the designer must have a full control over data representation (often a few contiguous bytes suffice where a conventional CAS would use up many scattered words); over sort and arithmetic algorithms; etc. The Gulo framework (as was the case with BEAR) allows one to separate the algebraic logic and the data representation, without blocking way for further optimizations. The new version is rather more efficient with very large expressions. Within its design requirements, it appears to be optimal and to cooperate nicely with the OS file caching mechanics, thus precluding a potential bottleneck.

The second software component deals with the solving of systems of linear equations that emerge in the IBP algorithms, including the generalized case discussed here. Zipper2 is a new version of Zipper that was used in the calculations of [18]. It is basically a solver of large systems of homogeneous linear equations with modifications. As mentioned above, the starting point was the Gauss type algorithms that we optimize based on statistics and a study of the algorithms formulated in terms of higher-level concepts (Gröbner bases etc.). The combination Gulo+Zipper2 is rather more efficient on bigger problems than the old one, BEAR+Zipper, even though the old version was 8 times faster in tests conducted within the project [18] than a competing software written in C++.

The Gauss-type algorithms contain two phases, a forward pass and a backward pass. The higher-level mathematical methods add a level of organization to these passes by using information about specific algebraic structures in the original problem. We work at a lower level and employ optimizations, so the separation into two phases is somewhat blurred. As a rule of thumb, both phases are similarly cumbersome, with a serious blow-up occurring around the middle. However, we found a trick, which we call R-trick, that essentially eliminates the second phase by relinquishing the task of construction of a general solution (another potential inhibition among many in this problem), with a significant speed gain. Since we work at a lower level than the algebra of polynomials, the R-trick is formulated accordingly, and we have not found a purely mathematical interpretation for it yet.

The second important mathematical optimization concerns the arithmetic of numerical coefficients. There is a wide room for experimentation here. (Note that such experimentation is hard if not impossible with C++, because it requires playing with data definitions freely, which would be an agony without all the safety features of Oberon.

Lastly, we have barely touched the options for storage optimization that are available in Gulo and Zipper2; significant optimization resources remain untapped.

### Partial *D*-operators

The third optimization is to drop the idea of constructing general *D*-operators that lower the powers of all polynomials in the product simultaneously as in eq. (2), in favour of the so-called partial operators that affect the power of only one factor:

$$\begin{aligned} &D_0(x,\mu,\partial)P_0^{\mu_0}P_1^{\mu_1}=b_0(\mu)P_0^{\mu_0-1}P_1^{\mu_1}, \quad b_0(\mu)\neq 0\\ &D_1(x,\mu,\partial)P_0^{\mu_0}P_1^{\mu_1}=b_1(\mu)P_0^{\mu_0}P_1^{\mu_1-1}, \quad b_1(\mu)\neq 0 \end{aligned} \tag{5}$$

Two such partial operators can be composed into a complete one that satisfies (2):

$$D(x,\mu,\partial) = D_1(x,\mu - e_0,\partial)D_0(x,\mu,\partial), \quad b(\mu) = b_0(\mu)b_1(\mu - e_0) \tag{6}$$

The point is, the system of equations for a partial operator does not look simpler than the one for the complete operator. Moreover, one does not a priori expect that the simplest complete operators (i.e. the ones with lowest degrees in $x$ and $\partial$) can be represented as compositions of partial ones. But the case turns out to be exactly that: the partial operators look in regard of overall complexity like factors of the complete ones and are found correspondingly faster. On top of that, the loop integrands in models with non-scalar particles have the two polynomials raised to different powers in different terms, so it is advantageous to raise the power of only one polynomial if the corresponding *D*-operator is simpler.

## First examples of 2-loop *D*-operators

The described optimizations at the three levels resulted in a speedup by several orders of magnitude compared with the older versions based on the software (BEAR v.2 + Zipper) that was by no means slow given the calculations done with it. The net effect is that it has proved possible to construct *D*-operators for the first true 2-loop integrals, and to do it rather efficiently. The simplest pair of polynomials corresponds to the "sunset" topology (cf. the 2-loop self-energy of the model $g\varphi^4$):

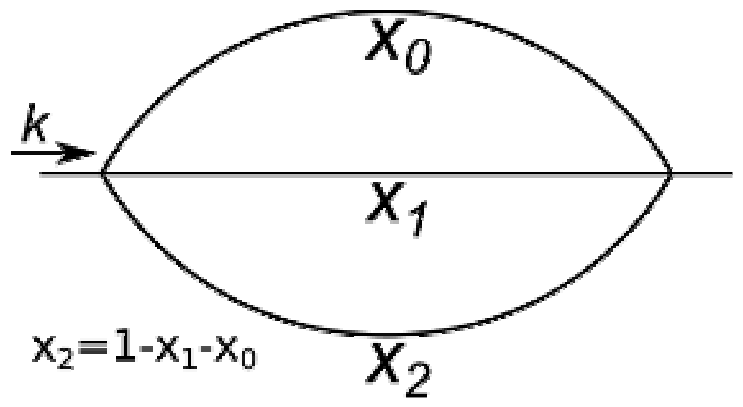

$$P_0 = (1 - x_0 - x_1)(x_0 + x_1) + x_0 x_1$$

$$P_1 = \left(m_0^2 x_0 + m_1^2 x_1 + m_2^2(1 - x_0 - x_1)\right)P_0 - k^2 x_0 x_1 (1 - x_0 - x_1)$$

Partial operators for this pair of polynomials with fixed masses and the external momentum squared are constructed in about one second on a low-end notebook. An example of such operators is shown in Appendix 1: the results are, of course, not meant to be read by a human. But with such speed, interesting games become possible. For instance, it proved possible to restore an exact analytical dependence of the *D*-operators on one of the kinematic parameters. Appendix 2 shows a partial operator for the sunset topology with the dependence on $k^2$ restored for the case of three equal masses. Such answers are easily verified by a direct substitution into eq. (2) within a general purpose CAS such as Wolfram Mathematica. Further examples can be found at [20].

## Conclusions

A surprising number of substantial optimizations have come up in this extremely cumbersome problem. As a result, the efficiency increased by several orders of magnitude compared with a previous version based on the software that was already rather more efficient than general purpose CASes. This paves way to problems of the next stage: the construction of *D*-operators for 2-loop integrals with more complex topologies (with a larger number of variables $x_i$ ), a systematic integrand simplification based on such *D*-operators, and a design of dedicated integration routines. One would, by simple extrapolation, expect more clever tricks to be discovered. In particular, the disk storage optimization options already available have practically not been employed yet. They are most effective for larger problems, and this appears to be a significant reserve.

## Appendix 1

Example of partial $D$-operators for the 2-loop “sunset” topology for

$m_0^2 = 1;\ m_1^2 = 2;\ m_2^2 = 3;\ k^2 = 1$.

The two polynomials are as follows:

$$P_0 = x_0 - x_0^2 + x_1 - x_0 x_1 - x_1^2$$

$$P_1 = 3x_0 - 5x_0^2 + 2x_0^3 + 3x_1 - 7x_0 x_1 + 4x_0^2 x_1 - 4x_1^2 + 4x_0 x_1^2 + x_1^3$$

The corresponding partial $D$-operators:

$$D_0 = C^{(0)} + A_0^{(0)}\partial_0 + A_1^{(0)}\partial_1 + F_{00}^{(0)}\partial_0^2 + F_{01}^{(0)}\partial_0\partial_1 + F_{11}^{(0)}\partial_1^2$$

$$b_0 = \mu_0\left(1/2 + \mu_0 + \mu_1\right)$$

where

$$C^{(0)} = -\frac{1101\mu_0}{92} + 93x_0\mu_0 - \frac{4779}{23}x_0^2\mu_0 + \frac{4455}{46}x_0^3\mu_0 + \frac{2577x_1\mu_0}{92} - \frac{12555}{92}x_0x_1\mu_0 + \frac{2673}{23}x_0^2x_1\mu_0$$
$$+\frac{405}{46}x_1^2\mu_0 + \frac{2673}{46}x_0x_1^2\mu_0 + \frac{891}{92}x_1^3\mu_0 - \frac{2765\mu_0^2}{276} + \frac{277}{3}x_0\mu_0^2 - \frac{14437}{69}x_0^2\mu_0^2 + \frac{8965}{92}x_0^3\mu_0^2 + \frac{6791}{276}x_1\mu_0^2$$
$$-\frac{73403}{552}x_0x_1\mu_0^2 + \frac{5379}{46}x_0^2x_1\mu_0^2 + \frac{1321}{92}x_1^2\mu_0^2 + \frac{5379}{92}x_0x_1^2\mu_0^2 + \frac{1793}{184}x_1^3\mu_0^2 - \frac{1377\mu_1}{92} + \frac{2403x_0\mu_1}{23}$$
$$-\frac{513}{2}x_0^2\mu_1 + \frac{6237}{46}x_0^3\mu_1 + \frac{3969x_1\mu_1}{92} - \frac{19467}{92}x_0x_1\mu_1 + \frac{8019}{46}x_0^2x_1\mu_1 - \frac{1593}{46}x_1^2\mu_1 + \frac{2673}{23}x_0x_1^2\mu_1$$
$$+\frac{891}{23}x_1^3\mu_1 - \frac{1678\mu_0\mu_1}{69} + \frac{13475}{69}x_0\mu_0\mu_1 - \frac{32162}{69}x_0^2\mu_0\mu_1 + \frac{21461}{92}x_0^3\mu_0\mu_1 + \frac{1496}{23}x_1\mu_0\mu_1$$
$$-\frac{189265}{552}x_0x_1\mu_0\mu_1 + \frac{26829}{92}x_0^2x_1\mu_0\mu_1 - \frac{2455}{138}x_1^2\mu_0\mu_1 + \frac{4026}{23}x_0x_1^2\mu_0\mu_1 + \frac{4477}{92}x_1^3\mu_0\mu_1 - \frac{1377\mu_1^2}{92}$$
$$+\frac{2403}{23}x_0\mu_1^2 - \frac{513}{2}x_0^2\mu_1^2 + + \frac{6237}{46}x_0^3\mu_1^2 + \frac{3969}{92}x_1\mu_1^2 - \frac{19467}{92}x_0x_1\mu_1^2 + \frac{8019}{46}x_0^2x_1\mu_1^2 - \frac{1593}{46}x_1^2\mu_1^2$$
$$+\frac{2673}{23}x_0x_1^2\mu_1^2 + \frac{891}{23}x_1^3\mu_1^2$$

$$A_0^{(0)} = -\frac{1}{4} + \frac{315x_0}{23} - \frac{10525x_0^2}{92} + \frac{7317x_0^3}{46} - \frac{2673x_0^4}{46} + \frac{278x_1}{23} - \frac{3001x_0x_1}{23} + \frac{9855}{46}x_0^2x_1 - \frac{4455}{46}x_0^3x_1$$
$$-\frac{5143x_1^2}{92} + \frac{3348}{23}x_0x_1^2 - \frac{8019}{92}x_0^2x_1^2 + 54x_1^3 - \frac{4455}{92}x_0x_1^3 - \frac{891x_1^4}{92} + \mu_0 + \frac{602x_0\mu_0}{69} - \frac{31019}{276}x_0^2\mu_0$$
$$+\frac{3683}{23}x_0^3\mu_0 - \frac{5379}{92}x_0^4\mu_0 + \frac{211x_1\mu_0}{23} - \frac{17405}{138}x_0x_1\mu_0 + \frac{58897}{276}x_0^2x_1\mu_0 - \frac{8965}{92}x_0^3x_1\mu_0 - \frac{29401}{552}x_1^2\mu_0$$
$$+\frac{39665}{276}x_0x_1^2\mu_0 - \frac{16137}{184}x_0^2x_1^2\mu_0 + \frac{163}{3}x_1^3\mu_0 - \frac{8965}{184}x_0x_1^3\mu_0 - \frac{1793}{184}x_1^4\mu_0 + \frac{1053x_0\mu_1}{92} - \frac{10341}{92}x_0^2\mu_1$$
$$+\frac{7317}{46}x_0^3\mu_1 - \frac{2673}{46}x_0^4\mu_1 + \frac{324x_1\mu_1}{23} - \frac{3024}{23}x_0x_1\mu_1 + \frac{9855}{46}x_0^2x_1\mu_1 - \frac{4455}{46}x_0^3x_1\mu_1 - \frac{5373}{92}x_1^2\mu_1$$
$$+\frac{3348}{23}x_0x_1^2\mu_1 - \frac{8019}{92}x_0^2x_1^2\mu_1 + 54x_1^3\mu_1 - \frac{4455}{92}x_0x_1^3\mu_1 - \frac{891}{92}x_1^4\mu_1$$

$$A_1^{(0)} = -\frac{3}{4} + \frac{81x_0}{23} + \frac{698x_0^2}{23} - \frac{2403x_0^3}{46} + \frac{891x_0^4}{46} + \frac{95x_1}{23} + \frac{737x_0x_1}{23} - \frac{2079}{92}x_0^2x_1 + \frac{656x_1^2}{23} - \frac{864}{23}x_0x_1^2$$
$$-\frac{2889x_1^3}{92} + \frac{303x_0\mu_0}{92} + \frac{8495}{276}x_0^2\mu_0 - \frac{14507}{276}x_0^3\mu_0 + \frac{1793}{92}x_0^4\mu_0 + \frac{509x_1\mu_0}{276} + \frac{9569}{276}x_0x_1\mu_0$$
$$-\frac{12505}{552}x_0^2x_1\mu_0 + \frac{17389}{552}x_1^2\mu_0 - \frac{5515}{138}x_0x_1^2\mu_0 - \frac{18959}{552}x_1^3\mu_0 + \frac{81x_0\mu_1}{23} + \frac{675}{23}x_0^2\mu_1 - \frac{2403}{46}x_0^3\mu_1$$
$$+\frac{891}{46}x_0^4\mu_1 + \frac{81x_1\mu_1}{92} + \frac{783}{23}x_0x_1\mu_1 - \frac{2079}{92}x_0^2x_1\mu_1 + \frac{702}{23}x_1^2\mu_1 - \frac{864}{23}x_0x_1^2\mu_1 - \frac{2889}{92}x_1^3\mu_1$$

$$F_{00}^{(0)} = -\frac{3}{4} + \frac{5x_0}{2} - \frac{11x_0^2}{4} + x_0^3 + \frac{7x_1}{2} - \frac{15x_0x_1}{2} + 4x_0^2x_1 - \frac{19x_1^2}{4} + \frac{19}{4}x_0x_1^2 + \frac{3x_1^3}{2}$$

$$F_{01}^{(0)} = \frac{3}{2} - \frac{13x_0}{2} + \frac{17x_0^2}{2} - \frac{7x_0^3}{2} - 6x_1 + \frac{33x_0x_1}{2} - 10x_0^2x_1 + \frac{29x_1^2}{4} - \frac{17}{2}x_0x_1^2 - \frac{11x_1^3}{4}$$

$$F_{11}^{(0)} = -\frac{3}{4} + 4x_0 - \frac{13x_0^2}{2} + 3x_0^3 + \frac{5x_1}{2} - 8x_0x_1 + \frac{11}{2}x_0^2x_1 - \frac{11x_1^2}{4} + 4x_0x_1^2 + x_1^3$$

$$D_1 = C^{(1)} + A_0^{(1)}\partial_0 + A_1^{(1)}\partial_1 + F_{00}^{(1)}\partial_0^2 + F_{01}^{(1)}\partial_0\partial_1 + F_{11}^{(1)}\partial_1^2$$

$$b_1 = \mu_1(1/2 + \mu_0 + \mu_1)$$

$$
\begin{aligned}
C^{(1)} = {} & \frac{207893\mu_0}{19044} - \frac{572641x_0\mu_0}{9522} + \frac{2073337x_0^2\mu_0}{19044} - \frac{578435x_0^3\mu_0}{12696} - \frac{938881x_1\mu_0}{38088} + \frac{5915389x_0x_1\mu_0}{76176} \\
& - \frac{115687x_0^2x_1\mu_0}{2116} + \frac{6963x_1^2\mu_0}{4232} - \frac{115687x_0x_1^2\mu_0}{4232} - \frac{115687x_1^3\mu_0}{25392} + \frac{54278\mu_0^2}{4761} - \frac{40033}{828}x_0\mu_0^2 \\
& + \frac{412783x_0^2\mu_0^2}{4761} - \frac{464915x_0^3\mu_0^2}{12696} - \frac{514643x_1\mu_0^2}{19044} + \frac{5246797x_0x_1\mu_0^2}{76176} - \frac{92983x_0^2x_1\mu_0^2}{2116} + \frac{42329x_1^2\mu_0^2}{12696} \\
& - \frac{92983x_0x_1^2\mu_0^2}{4232} - \frac{92983x_1^3\mu_0^2}{25392} + \frac{240577\mu_1}{19044} - \frac{332702x_0\mu_1}{4761} + \frac{652553x_0^2\mu_1}{4761} - \frac{809809x_0^3\mu_1}{12696} \\
& - \frac{437765x_1\mu_1}{12696} + \frac{9065585x_0x_1\mu_1}{76176} - \frac{347061x_0^2x_1\mu_1}{4232} + \frac{463985x_1^2\mu_1}{19044} - \frac{115687x_0x_1^2\mu_1}{2116} - \frac{115687x_1^3\mu_1}{6348} \\
& + \frac{472151\mu_0\mu_1}{19044} - \frac{644747x_0\mu_0\mu_1}{6348} + \frac{3533383x_0^2\mu_0\mu_1}{19044} - \frac{1056341x_0^3\mu_0\mu_1}{12696} - \frac{600173x_1\mu_0\mu_1}{9522} \\
& + \frac{13011877x_0x_1\mu_0\mu_1}{76176} - \frac{441133x_0^2x_1\mu_0\mu_1}{4232} + \frac{292945x_1^2\mu_0\mu_1}{9522} - \frac{133529x_0x_1^2\mu_0\mu_1}{2116} - \frac{9438}{529}x_1^3\mu_0\mu_1 \\
& + \frac{259271\mu_1^2}{19044} - \frac{256165x_0\mu_1^2}{4761} + \frac{19940}{207}x_0^2\mu_1^2 - \frac{141911x_0^3\mu_1^2}{3174} - \frac{59003x_1\mu_1^2}{1587} + \frac{1896949x_0x_1\mu_1^2}{19044} \\
& - \frac{60819x_0^2x_1\mu_1^2}{1058} + \frac{259847x_1^2\mu_1^2}{9522} - \frac{20273}{529}x_0x_1^2\mu_1^2 - \frac{20273x_1^3\mu_1^2}{1587}
\end{aligned}
$$

$$
\begin{aligned}
A_0^{(1)} = {} & -\frac{277}{2116} - \frac{426877x_0}{38088} + \frac{1278227x_0^2}{19044} - \frac{173457x_0^3}{2116} + \frac{115687x_0^4}{4232} - \frac{7557x_1}{1058} + \frac{337423x_0x_1}{4761} \\
& - \frac{4147223x_0^2x_1}{38088} + \frac{578435x_0^3x_1}{12696} + \frac{2086109x_1^2}{76176} - \frac{2734663x_0x_1^2}{38088} + \frac{347061x_0^2x_1^2}{8464} - \frac{972217x_1^3}{38088} \\
& + \frac{578435x_0x_1^3}{25392} + \frac{115687x_1^4}{25392} + \frac{1307\mu_0}{2116} - \frac{257167x_0\mu_0}{38088} + \frac{826103x_0^2\mu_0}{19044} - \frac{64389x_0^3\mu_0}{1058} + \frac{92983x_0^4\mu_0}{4232} \\
& - \frac{69541x_1\mu_0}{12696} + \frac{2117627x_0x_1\mu_0}{38088} - \frac{3288455x_0^2x_1\mu_0}{38088} + \frac{464915x_0^3x_1\mu_0}{12696} + \frac{1751183x_1^2\mu_0}{76176} \\
& - \frac{2155717x_0x_1^2\mu_0}{38088} + \frac{278949x_0^2x_1^2\mu_0}{8464} - \frac{764905x_1^3\mu_0}{38088} + \frac{464915x_0x_1^3\mu_0}{25392} + \frac{92983x_1^4\mu_0}{25392} + \frac{6037\mu_1}{6348} \\
& - \frac{161461x_0\mu_1}{19044} + \frac{707617x_0^2\mu_1}{19044} - \frac{326507x_0^3\mu_1}{6348} + \frac{20273x_0^4\mu_1}{1058} - \frac{104965x_1\mu_1}{12696} + \frac{1976005x_0x_1\mu_1}{38088} \\
& - \frac{720163x_0^2x_1\mu_1}{9522} + \frac{101365x_0^3x_1\mu_1}{3174} + \frac{900575x_1^2\mu_1}{38088} - \frac{958735x_0x_1^2\mu_1}{19044} + \frac{60819x_0^2x_1^2\mu_1}{2116} - \frac{324955x_1^3\mu_1}{19044} \\
& + \frac{101365x_0x_1^3\mu_1}{6348} + \frac{20273x_1^4\mu_1}{6348}
\end{aligned}
$$

$$A_1^{(1)} = \frac{1141}{1058} - \frac{50687x_0}{12696} - \frac{531373x_0^2}{38088} + \frac{965317x_0^3}{38088} - \frac{115687x_0^4}{12696} - \frac{119135x_1}{19044} - \frac{246223x_0x_1}{38088}$$
$$+ \frac{441395x_0^2x_1}{76176} - \frac{539249x_1^2}{76176} + \frac{216737x_0x_1^2}{19044} + \frac{913213x_1^3}{76176} - \frac{49\mu_0}{46} - \frac{32983x_0\mu_0}{12696} - \frac{264019x_0^2\mu_0}{38088}$$
$$+ \frac{706777x_0^3\mu_0}{38088} - \frac{92983x_0^4\mu_0}{12696} - \frac{48937x_1\mu_0}{9522} - \frac{86923x_0x_1\mu_0}{38088} + \frac{231683x_0^2x_1\mu_0}{76176} - \frac{186605x_1^2\mu_0}{76176}$$
$$+ \frac{92315x_0x_1^2\mu_0}{19044} + \frac{641581x_1^3\mu_0}{76176} - \frac{49\mu_1}{46} - \frac{1773x_0\mu_1}{1058} - \frac{45575x_0^2\mu_1}{9522} + \frac{74044x_0^3\mu_1}{4761} - \frac{20273x_0^4\mu_1}{3174}$$
$$- \frac{70159x_1\mu_1}{19044} - \frac{8467x_0x_1\mu_1}{19044} + \frac{26359x_0^2x_1\mu_1}{19044} - \frac{5423x_1^2\mu_1}{9522} + \frac{9599x_0x_1^2\mu_1}{9522} + \frac{87353x_1^3\mu_1}{19044}$$

$$F_{00}^{(1)} = \frac{1}{4} - \frac{3238x_0}{1587} - \frac{271x_0^2}{3174} + \frac{1930x_0^3}{529} - \frac{3751x_0^4}{2116} - \frac{26215x_1}{12696} + \frac{2843x_0x_1}{1058} + \frac{17971x_0^2x_1}{12696}$$
$$- \frac{1078}{529}x_0^3x_1 + \frac{9475x_1^2}{4232} - \frac{1641x_0x_1^2}{2116} - \frac{6193x_0^2x_1^2}{4232} - \frac{43x_1^3}{552} - \frac{2541x_0x_1^3}{4232} - \frac{407x_1^4}{4232}$$

$$F_{01}^{(1)} = -\frac{1}{2} + \frac{6530x_0}{1587} - \frac{29315x_0^2}{6348} + \frac{99x_0^3}{1058} + \frac{1925x_0^4}{2116} + \frac{13147x_1}{4232} - \frac{50869x_0x_1}{6348} + \frac{22127x_0^2x_1}{4232}$$
$$- \frac{33343x_1^2}{12696} + \frac{1727}{529}x_0x_1^2 + \frac{125x_1^3}{6348}$$

$$F_{11}^{(1)} = \frac{1}{4} - \frac{449x_0}{276} + \frac{281x_0^2}{92} - \frac{82x_0^3}{69} - \frac{11x_0^4}{46} - \frac{41x_1}{69} + \frac{781x_0x_1}{276} - \frac{235}{92}x_0^2x_1 + \frac{19x_1^2}{92} - \frac{83}{69}x_0x_1^2 + \frac{19x_1^3}{138}$$

## Appendix 2

Example of a partial $D$-operator for the 2-loop "sunset" topology with the exact dependence on the external momentum restored for the case of three equal non-zero masses (verified in Wolfram Mathematica). The two polynomials with the dependence on the dimensionless $\kappa^2 = k^2/m^2$ explicitly shown are as follows:

$$P_0 = (1 - x_0 - x_1)(x_0 + x_1) + x_0x_1$$
$$P_1 = (1 - x_0 - x_1)(x_0 + x_1) + x_0x_1 - \kappa^2 x_0x_1(1 - x_0 - x_1)$$

The partial operator $D_1$ and the corresponding polynomial $b_1$ are:

$$D_1 = C + A_0\partial_0 + A_1\partial_1 + (\kappa^2 - 1)(\kappa^2 - 9)(F_{00}\partial_0^2 + F_{01}\partial_0\partial_1 + F_{11}\partial_1^2)$$
$$b_1 = 4\kappa^2(\kappa^2 - 1)(\kappa^2 - 9)\mu_1(1 + 2\mu_0 + 2\mu_1)$$

where:

$$
\begin{aligned}
C = &\left(1944-2007\kappa^2+145\kappa^4+27\kappa^6-5\kappa^8\right)\mu_0 \\
&+\left(-5103+4212\kappa^2+774\kappa^4-210\kappa^6+13\kappa^8+2\kappa^{10}\right)x_0\mu_0 \\
&+\left(3645-1458\kappa^2-2160\kappa^4+276\kappa^6-\kappa^8-6\kappa^{10}\right)x_0^2\mu_0 \\
&+\left(-1458\kappa^2+1458\kappa^4-90\kappa^6-2\kappa^8+4\kappa^{10}\right)x_0^3\mu_0 \\
&+\left(-4374+4428\kappa^2-258\kappa^4-44\kappa^6+8\kappa^8\right)x_1\mu_0 \\
&+\left(5832-2916\kappa^2-2700\kappa^4+264\kappa^6+4\kappa^8-4\kappa^{10}\right)x_0x_1\mu_0 \\
&+\left(-4374\kappa^2+4374\kappa^4-216\kappa^6-30\kappa^8+6\kappa^{10}\right)x_0^2x_1\mu_0 \\
&+\left(1458-1458\kappa^2+72\kappa^4+10\kappa^6-2\kappa^8\right)x_1^2\mu_0 \\
&+\left(-1458\kappa^2+1458\kappa^4-72\kappa^6-10\kappa^8+2\kappa^{10}\right)x_0x_1^2\mu_0 \\
&+\left(1944-1998\kappa^2+270\kappa^4-2\kappa^6-6\kappa^8\right)\mu_0^2 \\
&+\left(-5103+4374\kappa^2+216\kappa^4-130\kappa^6+15\kappa^8+4\kappa^{10}\right)x_0\mu_0^2 \\
&+\left(3645-1458\kappa^2-1944\kappa^4+374\kappa^6-13\kappa^8-12\kappa^{10}\right)x_0^2\mu_0^2 \\
&+\left(-1458\kappa^2+1458\kappa^4-198\kappa^6+14\kappa^8+8\kappa^{10}\right)x_0^3\mu_0^2 \\
&+\left(-4374+4374\kappa^2-486\kappa^4-6\kappa^6+12\kappa^8\right)x_1\mu_0^2 \\
&+\left(5832-2916\kappa^2-2268\kappa^4+332\kappa^6-12\kappa^8-8\kappa^{10}\right)x_0x_1\mu_0^2 \\
&+\left(-4374\kappa^2+4374\kappa^4-486\kappa^6-6\kappa^8+12\kappa^{10}\right)x_0^2x_1\mu_0^2 \\
&+\left(1458-1458\kappa^2+162\kappa^4+2\kappa^6-4\kappa^8\right)x_1^2\mu_0^2 \\
&+\left(-1458\kappa^2+1458\kappa^4-162\kappa^6-2\kappa^8+4\kappa^{10}\right)x_0x_1^2\mu_0^2 \\
&+\left(1944-1845\kappa^2-8\kappa^4+15\kappa^6-2\kappa^8\right)\mu_1 \\
&+\left(-5103+3969\kappa^2+1053\kappa^4-247\kappa^6+10\kappa^8+2\kappa^{10}\right)x_0\mu_1 \\
&+\left(3645-1944\kappa^2-1755\kappa^4+336\kappa^6-12\kappa^8-6\kappa^{10}\right)x_0^2\mu_1 \\
&+\left(-729\kappa^2+729\kappa^4-63\kappa^6+7\kappa^8+4\kappa^{10}\right)x_0^3\mu_1 \\
&+\left(-4374+3456\kappa^2+696\kappa^4-66\kappa^6-2\kappa^8+2\kappa^{10}\right)x_1\mu_1 \\
&+\left(5832-486\kappa^2-5130\kappa^4+378\kappa^6+22\kappa^8-8\kappa^{10}\right)x_0x_1\mu_1 \\
&+\left(-4374\kappa^2+4374\kappa^4-216\kappa^6-30\kappa^8+6\kappa^{10}\right)x_0^2x_1\mu_1 \\
&+\left(1458+972\kappa^2-2322\kappa^4+84\kappa^6+20\kappa^8-4\kappa^{10}\right)x_1^2\mu_1 \\
&+\left(-4374\kappa^2+4374\kappa^4-216\kappa^6-30\kappa^8+6\kappa^{10}\right)x_0x_1^2\mu_1 \\
&+\left(-1458\kappa^2+1458\kappa^4-72\kappa^6-10\kappa^8+2\kappa^{10}\right)x_1^3\mu_1 \\
&+\left(3888-3510\kappa^2+50\kappa^6-12\kappa^8\right)\mu_0\mu_1
\end{aligned}
$$

$$+\left(-10206+7938\kappa^2+1260\kappa^4-252\kappa^6-6\kappa^8+10\kappa^{10}\right)x_0\mu_0\mu_1$$
$$+\left(7290-2997\kappa^2-3573\kappa^4+363\kappa^6+67\kappa^8-30\kappa^{10}\right)x_0^2\mu_0\mu_1$$
$$+\left(-2187\kappa^2+2025\kappa^4-117\kappa^6-21\kappa^8+20\kappa^{10}\right)x_0^3\mu_0\mu_1$$
$$+\left(-8748+7290\kappa^2+594\kappa^4-222\kappa^6+26\kappa^8+4\kappa^{10}\right)x_1\mu_0\mu_1$$
$$+\left(11664-2754\kappa^2-7434\kappa^4+678\kappa^6+50\kappa^8-28\kappa^{10}\right)x_0x_1\mu_0\mu_1$$
$$+\left(-8748\kappa^2+8262\kappa^4-378\kappa^6-126\kappa^8+30\kappa^{10}\right)x_0^2x_1\mu_0\mu_1$$
$$+\left(2916-324\kappa^2-2304\kappa^4+300\kappa^6-4\kappa^8-8\kappa^{10}\right)x_1^2\mu_0\mu_1$$
$$+\left(-5832\kappa^2+5670\kappa^4-450\kappa^6-46\kappa^8+18\kappa^{10}\right)x_0x_1^2\mu_0\mu_1$$
$$+\left(-1458\kappa^2+1458\kappa^4-162\kappa^6-2\kappa^8+4\kappa^{10}\right)x_1^3\mu_0\mu_1$$
$$+\left(1944-1512\kappa^2-252\kappa^4+32\kappa^6-4\kappa^8\right)\mu_1^2$$
$$+\left(-5103+3564\kappa^2+981\kappa^4-52\kappa^6-28\kappa^8+6\kappa^{10}\right)x_0\mu_1^2$$
$$+\left(3645-1539\kappa^2-1647\kappa^4+9\kappa^6+78\kappa^8-18\kappa^{10}\right)x_0^2\mu_1^2$$
$$+\left(-729\kappa^2+648\kappa^4-9\kappa^6-26\kappa^8+12\kappa^{10}\right)x_0^3\mu_1^2$$
$$+\left(-4374+2916\kappa^2+936\kappa^4-38\kappa^6-22\kappa^8+6\kappa^{10}\right)x_1\mu_1^2$$
$$+\left(5832+162\kappa^2-4860\kappa^4-30\kappa^6+136\kappa^8-24\kappa^{10}\right)x_0x_1\mu_1^2$$
$$+\left(-4374\kappa^2+3888\kappa^4+108\kappa^6-120\kappa^8+18\kappa^{10}\right)x_0^2x_1\mu_1^2$$
$$+\left(1458+1134\kappa^2-2160\kappa^4-78\kappa^6+74\kappa^8-12\kappa^{10}\right)x_1^2\mu_1^2$$
$$+\left(-4374\kappa^2+3888\kappa^4+108\kappa^6-120\kappa^8+18\kappa^{10}\right)x_0x_1^2\mu_1^2$$
$$+\left(-1458\kappa^2+1296\kappa^4+36\kappa^6-40\kappa^8+6\kappa^{10}\right)x_1^3\mu_1^2$$

$$A_0=18-47\kappa^2+32\kappa^4-3\kappa^6$$
$$+\left(-1755+1689\kappa^2+5\kappa^4-45\kappa^6+6\kappa^8\right)x_0$$
$$+\left(3888-2988\kappa^2-841\kappa^4+196\kappa^6-9\kappa^8-2\kappa^{10}\right)x_0^2$$
$$+\left(-2187+729\kappa^2+1449\kappa^4-185\kappa^6+2\kappa^8+4\kappa^{10}\right)x_0^3$$
$$+\left(729\kappa^2-729\kappa^4+45\kappa^6+\kappa^8-2\kappa^{10}\right)x_0^4$$
$$+\left(-972+1026\kappa^2-84\kappa^4-22\kappa^6+4\kappa^8\right)x_1$$
$$+\left(5346-4446\kappa^2-670\kappa^4+112\kappa^6-4\kappa^8-2\kappa^{10}\right)x_0x_1$$
$$+\left(-4374+486\kappa^2+3708\kappa^4-256\kappa^6-18\kappa^8+6\kappa^{10}\right)x_0^2x_1$$

$$+\left(2916\kappa^2-2916\kappa^4+144\kappa^6+20\kappa^8-4\kappa^{10}\right)x_0^3x_1$$
$$+\left(2430-2466\kappa^2+136\kappa^4+34\kappa^6-6\kappa^8\right)x_1^2$$
$$+\left(-2916+486\kappa^2+2268\kappa^4-114\kappa^6-16\kappa^8+4\kappa^{10}\right)x_0x_1^2$$
$$+\left(2916\kappa^2-2916\kappa^4+144\kappa^6+20\kappa^8-4\kappa^{10}\right)x_0^2x_1^2$$
$$+\left(-1458+1458\kappa^2-72\kappa^4-10\kappa^6+2\kappa^8\right)x_1^3$$
$$+\left(1458\kappa^2-1458\kappa^4+72\kappa^6+10\kappa^8-2\kappa^{10}\right)x_0x_1^3$$
$$+\left(54\kappa^2-60\kappa^4+6\kappa^6\right)\mu_0$$
$$+\left(-1701+1548\kappa^2-70\kappa^4+20\kappa^6+3\kappa^8\right)x_0\mu_0$$
$$+\left(3888-3132\kappa^2-312\kappa^4+48\kappa^6-4\kappa^{10}\right)x_0^2\mu_0$$
$$+\left(-2187+729\kappa^2+1251\kappa^4-181\kappa^6+4\kappa^8+8\kappa^{10}\right)x_0^3\mu_0$$
$$+\left(729\kappa^2-729\kappa^4+99\kappa^6-7\kappa^8-4\kappa^{10}\right)x_0^4\mu_0$$
$$+\left(-972+936\kappa^2-56\kappa^4-8\kappa^6+4\kappa^8\right)x_1\mu_0$$
$$+\left(5346-4320\kappa^2-432\kappa^4+92\kappa^6-10\kappa^8-4\kappa^{10}\right)x_0x_1\mu_0$$
$$+\left(-4374+486\kappa^2+3366\kappa^4-386\kappa^6+12\kappa^{10}\right)x_0^2x_1\mu_0$$
$$+\left(2916\kappa^2-2916\kappa^4+324\kappa^6+4\kappa^8-8\kappa^{10}\right)x_0^3x_1\mu_0$$
$$+\left(2430-2430\kappa^2+258\kappa^4+6\kappa^6-8\kappa^8\right)x_1^2\mu_0$$
$$+\left(-2916+486\kappa^2+2106\kappa^4-262\kappa^6+2\kappa^8+8\kappa^{10}\right)x_0x_1^2\mu_0$$
$$+\left(2916\kappa^2-2916\kappa^4+324\kappa^6+4\kappa^8-8\kappa^{10}\right)x_0^2x_1^2\mu_0$$
$$+\left(-1458+1458\kappa^2-162\kappa^4-2\kappa^6+4\kappa^8\right)x_1^3\mu_0$$
$$+\left(1458\kappa^2-1458\kappa^4+162\kappa^6+2\kappa^8-4\kappa^{10}\right)x_0x_1^3\mu_0$$
$$+\left(90\kappa^2-100\kappa^4+10\kappa^6\right)\mu_1$$
$$+\left(-1701+1305\kappa^2+209\kappa^4-17\kappa^6+4\kappa^8\right)x_0\mu_1$$
$$+\left(3888-2754\kappa^2-597\kappa^4-78\kappa^6+35\kappa^8-6\kappa^{10}\right)x_0^2\mu_1$$
$$+\left(-2187+486\kappa^2+1296\kappa^4+78\kappa^6-61\kappa^8+12\kappa^{10}\right)x_0^3\mu_1$$
$$+\left(729\kappa^2-648\kappa^4-9\kappa^6+22\kappa^8-6\kappa^{10}\right)x_0^4\mu_1$$
$$+\left(-972+720\kappa^2+184\kappa^4-32\kappa^6+4\kappa^8\right)x_1\mu_1$$
$$+\left(5346-3618\kappa^2-1086\kappa^4+12\kappa^6+24\kappa^8-6\kappa^{10}\right)x_0x_1\mu_1$$
$$+\left(-4374+3456\kappa^4+114\kappa^6-110\kappa^8+18\kappa^{10}\right)x_0^2x_1\mu_1$$
$$+\left(2916\kappa^2-2592\kappa^4-72\kappa^6+80\kappa^8-12\kappa^{10}\right)x_0^3x_1\mu_1$$
$$+\left(2430-2052\kappa^2-180\kappa^4+68\kappa^6-10\kappa^8\right)x_1^2\mu_1$$

$$+\left(-2916+162\kappa^2+2178\kappa^4+58\kappa^6-70\kappa^8+12\kappa^{10}\right)x_0x_1^2\mu_1$$
$$+\left(2916\kappa^2-2592\kappa^4-72\kappa^6+80\kappa^8-12\kappa^{10}\right)x_0^2x_1^2\mu_1$$
$$+\left(-1458+1296\kappa^2+36\kappa^4-40\kappa^6+6\kappa^8\right)x_1^3\mu_1$$
$$+\left(1458\kappa^2-1296\kappa^4-36\kappa^6+40\kappa^8-6\kappa^{10}\right)x_0x_1^3\mu_1$$

$$A_1=18-11\kappa^2-8\kappa^4+\kappa^6$$
$$+\left(-243+351\kappa^2-117\kappa^4+5\kappa^6\right)x_0$$
$$+\left(-486+333\kappa^2+140\kappa^4-19\kappa^6\right)x_0^2$$
$$+\left(729-729\kappa^2+27\kappa^4+9\kappa^6\right)x_0^3$$
$$+\left(-1026+1032\kappa^2-57\kappa^4-6\kappa^6+\kappa^8\right)x_1$$
$$+\left(972-801\kappa^2-175\kappa^4+69\kappa^6-5\kappa^8\right)x_0x_1$$
$$+\left(486\kappa^2-405\kappa^4-60\kappa^6+11\kappa^8\right)x_0^2x_1$$
$$+\left(-729\kappa^2+729\kappa^4-27\kappa^6-9\kappa^8\right)x_0^3x_1$$
$$+\left(972-981\kappa^2+61\kappa^4+5\kappa^6-\kappa^8\right)x_1^2$$
$$+\left(-729+486\kappa^2+252\kappa^4-70\kappa^6+5\kappa^8\right)x_0x_1^2$$
$$+\left(-36\kappa^2+40\kappa^4-4\kappa^6\right)\mu_0$$
$$+\left(-243+423\kappa^2-179\kappa^4-11\kappa^6+2\kappa^8\right)x_0\mu_0$$
$$+\left(-486+378\kappa^2+50\kappa^6-6\kappa^8\right)x_0^2\mu_0$$
$$+\left(729-729\kappa^2+99\kappa^4-31\kappa^6+4\kappa^8\right)x_0^3\mu_0$$
$$+\left(-972+1008\kappa^2-154\kappa^4+4\kappa^6+2\kappa^8\right)x_1\mu_0$$
$$+\left(972-945\kappa^2+57\kappa^4+41\kappa^6-5\kappa^8\right)x_0x_1\mu_0$$
$$+\left(486\kappa^2-396\kappa^4-34\kappa^6+8\kappa^8\right)x_0^2x_1\mu_0$$
$$+\left(-729\kappa^2+729\kappa^4-63\kappa^6-9\kappa^8\right)x_0^3x_1\mu_0$$
$$+\left(972-972\kappa^2+114\kappa^4-2\kappa^8\right)x_1^2\mu_0$$
$$+\left(-729+486\kappa^2+162\kappa^4-34\kappa^6+3\kappa^8\right)x_0x_1^2\mu_0$$
$$+\left(-243+288\kappa^2-29\kappa^4-26\kappa^6+2\kappa^8\right)x_0\mu_1$$
$$+\left(-486+432\kappa^2-60\kappa^4+56\kappa^6-6\kappa^8\right)x_0^2\mu_1$$
$$+\left(729-648\kappa^2+9\kappa^4-22\kappa^6+4\kappa^8\right)x_0^3\mu_1$$
$$+\left(-972+846\kappa^2+44\kappa^4-34\kappa^6+4\kappa^8\right)x_1\mu_1$$

$$+\left(972-729\kappa^2-192\kappa^4+75\kappa^6-6\kappa^8\right)x_0x_1\mu_1$$
$$+\left(486\kappa^2-378\kappa^4-54\kappa^6+10\kappa^8\right)x_0^2x_1\mu_1$$
$$+\left(-729\kappa^2+648\kappa^4+27\kappa^6-18\kappa^8\right)x_0^3x_1\mu_1$$
$$+\left(972-918\kappa^2+36\kappa^4+26\kappa^6-4\kappa^8\right)x_1^2\mu_1$$
$$+\left(-729+405\kappa^2+261\kappa^4-53\kappa^6+4\kappa^8\right)x_0x_1^2\mu_1$$

$$F_{00}=2-4x_0-7\kappa^2x_0+2x_0^2+6\kappa^2x_0^2+3\kappa^4x_0^2+\kappa^2x_0^3-6\kappa^4x_0^3+3\kappa^4x_0^4$$
$$-8x_1-2\kappa^2x_1+8x_0x_1+14\kappa^2x_0x_1+2\kappa^4x_0x_1-12\kappa^2x_0^2x_1-6\kappa^4x_0^2x_1$$
$$+4\kappa^4x_0^3x_1+8x_1^2+2\kappa^2x_1^2-10\kappa^2x_0x_1^2-2\kappa^4x_0x_1^2+2\kappa^4x_0^2x_1^2$$

$$F_{01}=-4+12x_0+5\kappa^2x_0-8x_0^2-6\kappa^2x_0^2-2\kappa^4x_0^2+\kappa^2x_0^3+4\kappa^4x_0^3-2\kappa^4x_0^4$$
$$+12x_1-20x_0x_1-8\kappa^2x_0x_1-\kappa^4x_0x_1+16\kappa^2x_0^2x_1+2\kappa^4x_0^2x_1-\kappa^4x_0^3x_1-8x_1^2$$
$$+7\kappa^2x_0x_1^2+\kappa^4x_0x_1^2$$

$$F_{11}=2-8x_0+8x_0^2-4x_1+8x_0x_1+4\kappa^2x_0x_1-8\kappa^2x_0^2x_1+2x_1^2-4\kappa^2x_0x_1^2$$